\documentclass{article}
\usepackage[T1]{fontenc}
\usepackage{amssymb}
\usepackage{amsfonts}
\usepackage{amsmath}
\usepackage{epsfig}
\usepackage{graphicx}
\title{An experimental test of the Jarzynski equality in a mechanical experiment}
\author{F. Douarche, S. Ciliberto,  A. Petrosyan I. Rabbiosi \\
     Laboratoire de Physique de l'ENS Lyon CNRS UMR 5672 \\
    46, All\'ee d'Italie - 69364 Lyon Cedex 07, France
}

\def\ga{\gamma}
\def\om{\omega}
\def\lam{\lambda}

\def\d{\,\mathrm{d}}
\begin{document}
\maketitle
\begin{abstract}
We have experimentally checked the Jarzynski equality and the Crooks relation
\cite{jarzynski1, crooks1, jarzynski2} on the thermal fluctuations of a
macroscopic mechanical oscillator in contact with a heat reservoir. We found
that, independently of the time scale and amplitude of the driving force, both
relations are satisfied. These results give credit, at least in the case of
Gaussian fluctuations, to the use of these relations in biological and chemical
systems to estimate the free energy difference $\Delta F$ between two
equilibrium states \cite{ritort, bustamante}. An alternative method to estimate
$\Delta F$ in isothermal process is proposed too.
\end{abstract}
\section{Introduction}
Many systems in Nature and in technological applications work out of
equilibrium. However, a precise estimation of the free energy difference
$\Delta F$ between two equilibrium states $A$ and $B$ of these systems is
extremely useful to increase our knowledge of the underlying physical processes
which control their dynamical behaviour. It is well known that $\Delta F$ can be
estimated by perturbing the system with an external parameter $\lam$ and by
measuring the work $W$ done to drive the system from $A$ to $B$. However this
method gives in general an overestimation of $\Delta F$ because
$W \geq \Delta F$, where the equality holds if and only if the perturbation is
infinitely slow. In other words the path $\ga$ to go from $A$ to $B$ has to be a
reversible one. In many systems, because of unavoidable experimental and
environmental constraints, the path $\ga$ is not a reversible one, that is the
system cannot be driven from $A$ to $B$ in a time much longer than its
relaxation time. This may happen for example in all the systems where thermal
fluctuations cannot be neglected and the external power injected into the system
is comparable to the thermal energy. In 1997 \cite{jarzynski1} Jarzynski derived
an equality which relates the free energy difference of a system in contact with
a heat reservoir to the probability distribution function (pdf) of the work
performed on the system to drive it from $A$ to $B$ along any path $\ga$ in the
system parameter space. Specifically, when $\lam$ is varied from time $t = 0$ to
$t = t_s$, Jarzynski defines for one realization of the ``switching process''
from $A$ to $B$ the work performed on the system as
\begin{equation}
    W = \int_{0}^{t_s} \dot{\lambda}\, \frac{\partial H_{\lam} [z(t)]}
    {\partial \lam} \d t,
    \label{work}
\end{equation}
\noindent where $z$ denotes the phase-space point of the system and $H_{\lam}$
its $\lam$-parametrized Hamiltonian (see also \cite{landau_stat}). One can
consider an ensemble of realizations of this ``switching process'' with initial
conditions all starting in the same initial equilibrium state. Then $W$ may be
computed for each trajectory in the ensemble. The Jarzynski equality (JR) states
that \cite{jarzynski1}
\begin{equation}
    \Delta F = -\frac{1}{\beta} \log \langle \exp{[-\beta W]} \rangle,
    \label{JE}
\end{equation}
where $\langle{\cdot}\rangle$ denotes the ensemble average, $\beta^{-1} = k_B T$
with $k_B$ the Boltzmann constant and $T$ the temperature. In other words
$\langle \exp{[-\beta W_{\mathrm{diss}}]} \rangle = 1$, since we can always
write $W = \Delta F+ W_{\mathrm{diss}}$ where $W_{\mathrm{diss}}$ is the
dissipated work. Thus it is easy to see that there must exist some paths $\ga$
such that $W_{\textrm{diss}} \leq 0$. Moreover, the inequality
$\langle \exp{x} \rangle \geq \exp{\langle x \rangle}$ allows us to recover the
second principle, namely $\langle W_{\textrm{diss}} \rangle \geq 0$, i.e.
$\langle W \rangle \geq \Delta F$. From an experimental point of view the JE is
quite useful because there is no restriction on the choice of the path $\ga$ and
it overcomes the above mentioned experimental difficulties. Numerous derivations
of the JE has been produced
\cite{crooks2, jarzynski3, mazonka_jarzynski, crooks3, wojcik}, but it seems
from the recent criticisms of Cohen and Mauzerall \cite{cohen} that this result
is still under debate. Without willing to enter in this letter into the
theoretical debate, we think that it is important to experimentally check
the JE on a very simple and controlled system in order to safely use it in more
complex cases as the biological and chemical ones, where it is much more
difficult to verify the results with other methods. For this reason we
experimentally probe a model system: a macroscopic mechanical oscillator driven
out of equilibrium, between two equilibrium states $A$ and $B$, by a small
external force. We show that the JE is experimentally accessible and valid, and
does not depend on the oscillator's damping, on the driving force's switching
rate and on its amplitude. In our experiment we can also check the Crooks
relation (CR) which is somehow related to the JE and which gives useful and
complementary information on the dissipated work. Crooks considers the forward
work $W_{\mathrm{f}}$ to drive the system from $A$ to $B$ and the backward work
$W_{\mathrm{b}}$ to drive it from $B$ to $A$. If the work pdfs during the
forward and backward processes are
$\mathrm{P}_{\mathrm{f}}(W)$ and $\mathrm{P}_{\mathrm{b}}(W)$, one has
\cite{crooks1, jarzynski2}
\begin{equation}
    \frac{\mathrm{P}_{\mathrm{f}}(W)}{\mathrm{P}_{\mathrm{b}}(-W)}
    = \exp{(\beta [W-\Delta F])}
    = \exp{[\beta W_{\mathrm{diss}}]}
    \label{crooks}.
\end{equation}
A simple calculation from Eq.\,(\ref{crooks}) leads to Eq.\,(\ref{JE}). However,
from an experimental point of view this relation is extremely useful because one
immediately sees that the crossing point of the two pdfs, that is the point
where $\mathrm{P}_{\mathrm{f}}(W) = \mathrm{P}_{\mathrm{b}}(-W)$, is precisely
$\Delta F$. Thus one has another mean to check the computed free energy by
looking at the pdfs crossing point $\Delta F_{\mathrm{\times}}$. Let us examine
in some detail the Gaussian case:
$\mathrm{P}(W) \propto \exp{\Bigl(-\frac{[W - \langle W \rangle]^2}{2 \sigma_W^2}}\Bigr)$
leads to $\Delta F = \langle W \rangle - \frac{\beta \sigma_W^2}{2}$, i.e.
$\langle W_{\textrm{diss}} \rangle = \frac{\beta \sigma_W^2}{2} > 0$.
Furthermore, it is easy to see from Eq.\,(\ref{crooks}) that if
$\mathrm{P}_{\mathrm{f}}(W)$ and $\mathrm{P}_{\mathrm{b}}(-W)$ are Gaussian,
then $\Delta F = \frac{\langle W \rangle_{\mathrm{f}} - \langle W \rangle_{\mathrm{b}}}{2}$
and $\beta \sigma_W^2 = \langle W \rangle_{\mathrm{f}} + \langle W \rangle_{\mathrm{b}} =
2\,\langle W_{\mathrm{diss}} \rangle$.
Thus in the case of Gaussian statistics $\Delta F$ and $W_{\mathrm{diss}}$ can
be computed by using just the mean values and the variance of the work $W$.

Before describing the experiment, we want to discuss several important points.
The first is the definition of the work given in Eq.\,(\ref{work}), which is not
the classical one. Let us consider, for example, that $\lambda$ is a mechanical
torque $M$ applied to a mechanical system $\Xi$, and
$-\partial H_{\lam} / \partial \lam$ the associated angular displacement
$\theta$. Then, from Eq.\,(\ref{work}), one has
\begin{equation}
    W = -\int_{0}^{t_s} \dot{M} \theta \d t = -\Bigl[M \theta \Bigr]_{0}^{t_s} - W^{\mathrm{cl}}
    \qquad\textrm{where}\qquad W^{\mathrm{cl}} = -\int_{0}^{t_s} M \dot{\theta} \d t
\end{equation}
is the classical work. Thus $W$ and $W^{\mathrm{cl}}$ are related but they are
not exactly the same and we will show that this makes an important difference in
the fluctuations of these two quantities. The second point concerns the
$\Delta F$ computed by the JE in the case of a driven system, composed by $\Xi$
plus the external driving. The total free energy difference
is
$\Delta F = \Delta F_0 - \Phi$ where $\Delta F_0$ is the free energy of $\Xi$
and $\Phi = \Bigl[M \theta \Bigr]_A^B$ the energy difference of the forcing.
The JE computes the $\Delta F$ of the driven system and not that of the system
alone  which is $\Delta F_0$. This is an important observation in view of all
applications where an external parameter is added to $\Xi$ in order to measure
$\Delta F_0$ \cite{bustamante}. Finally we point out that, in an isothermal
process,  $\Delta F_0$ can be easily computed, without using the JE and the CR,
if $W^{\mathrm{cl}}$ is Gaussian distributed with variance
$\sigma_{W^{\mathrm{cl}}}^2$. Indeed the crossing point
$W_{\times}^{\mathrm{cl}}$ of the two Gaussian pdfs
$\mathrm{P}_{\mathrm{f}} (W^{\mathrm{cl}})$ and
$\mathrm{P}_{\mathrm{b}} (-W^{\mathrm{cl}})$ is
\begin{equation}
    W_{\times}^{\mathrm{cl}} = \frac{\langle W^{\mathrm{cl}} \rangle_{\mathrm{f}}
    - \langle W^{\mathrm{cl}} \rangle_{\mathrm{b}}}{2},
\end{equation}
which by definition is just $-\Delta F_0$, i.e.
$ W_{\times}^{\mathrm{cl}}= -\Delta F_0$. Furthermore
$\langle W^{\mathrm{cl}} \rangle_{\mathrm{f}} + \langle W^{\mathrm{cl}}
\rangle_{\mathrm{b}} = -2\,\langle W_{\mathrm{diss}} \rangle$ by definition, but
in this case
$2\,\langle W_{\mathrm{diss}} \rangle \neq \beta \sigma_{W^{\mathrm{cl}}}^2$.
\begin{figure}
    \begin{center}
    \includegraphics[width=5cm, angle=0]{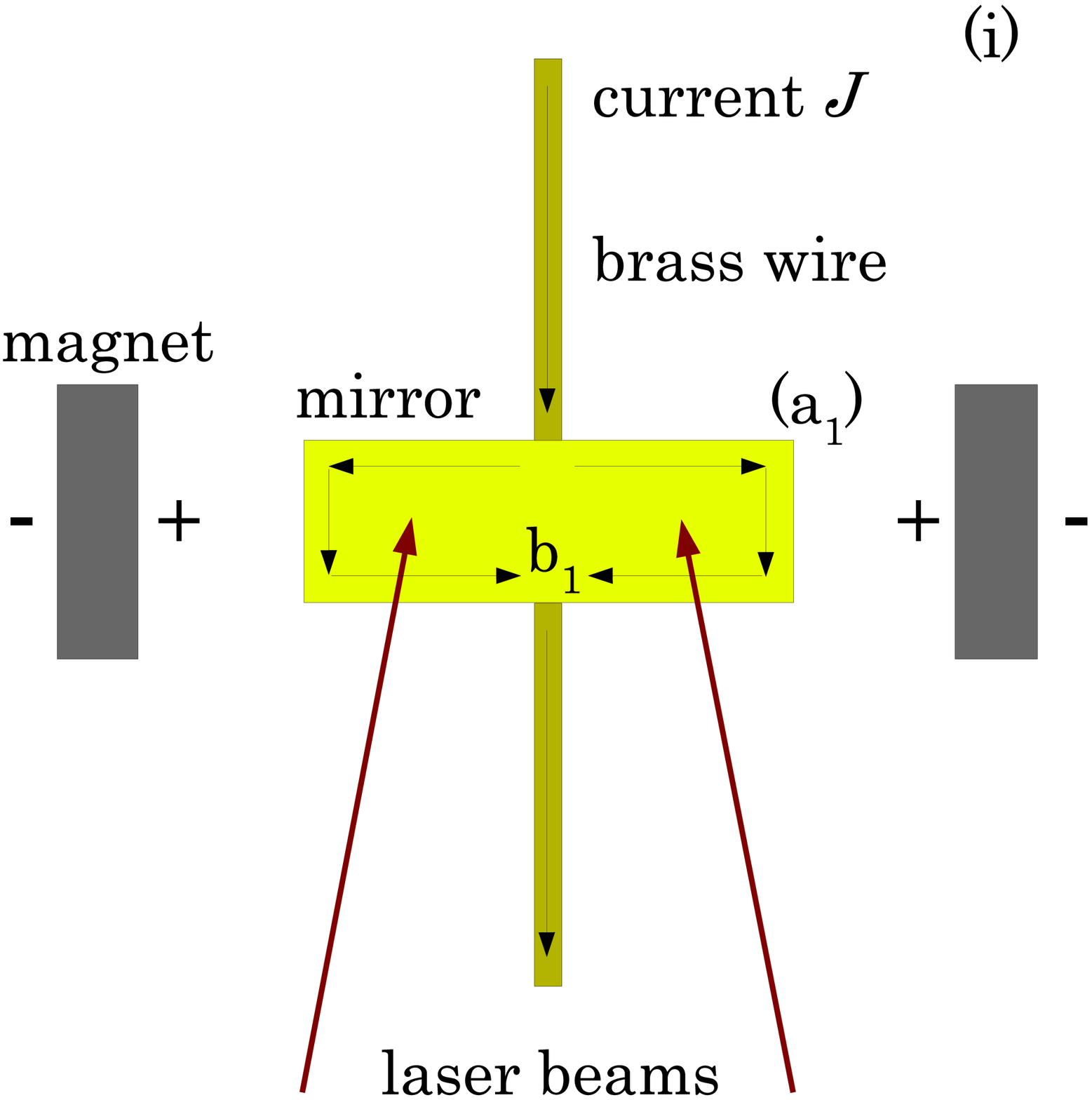}
    \includegraphics[width=6cm, angle=0]{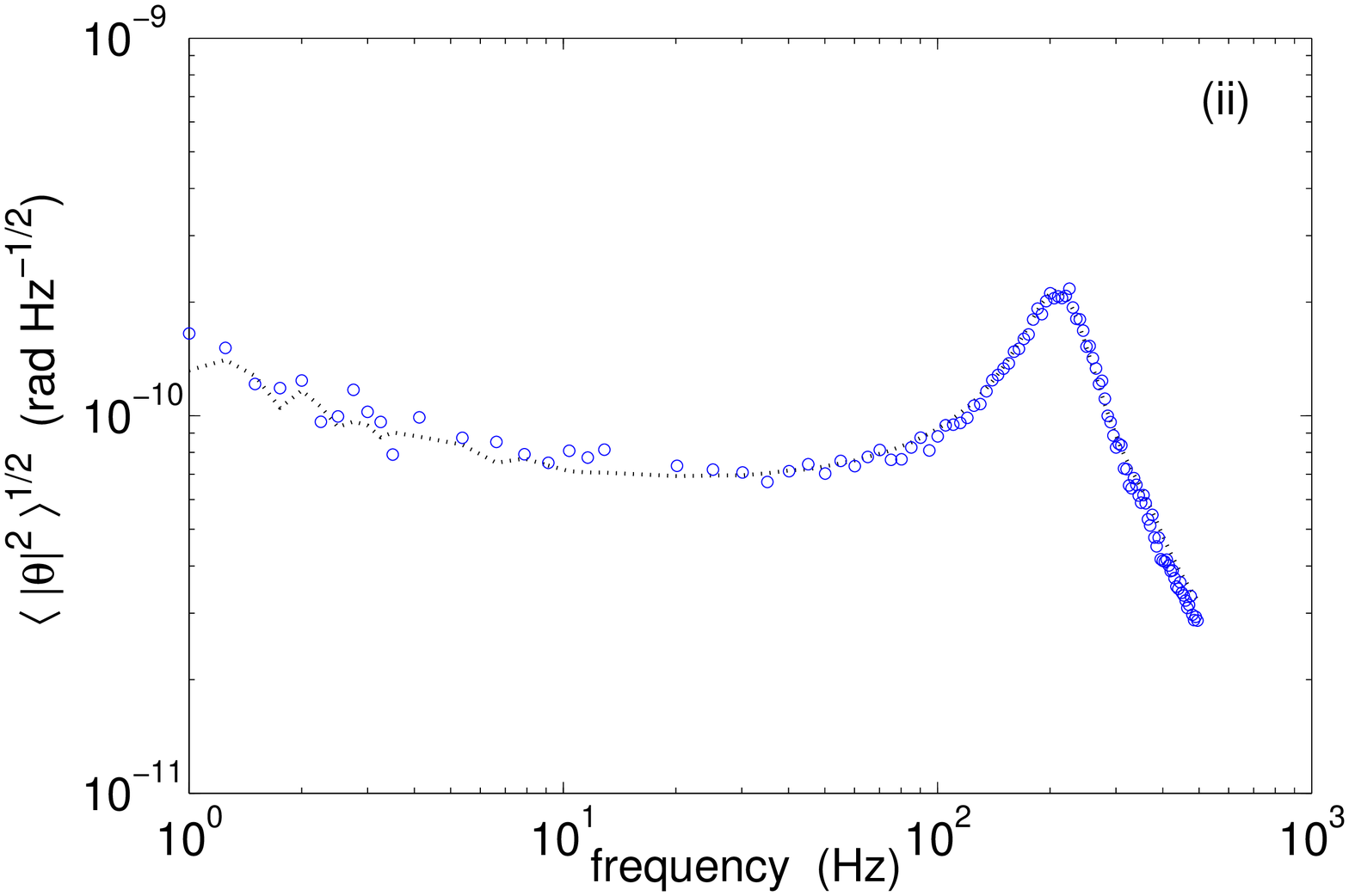}
    \caption{(i) Schematic drawing of the oscillator; (ii) FDT check in oil:
    the circles $(\circ)$ correspond to the direct measurement of the noise and
    the dashed curve is $\sqrt{\langle { \vert \hat{\theta} \vert }^2 \rangle}$
    computed by inserting the measured response function in the FDT, Eq.\,(\ref{fdt})}
    \label{setup_and_fdt_check}
    \end{center}
\end{figure}
\begin{figure}
    \begin{center}
    \includegraphics[width=4.6cm, angle=0]{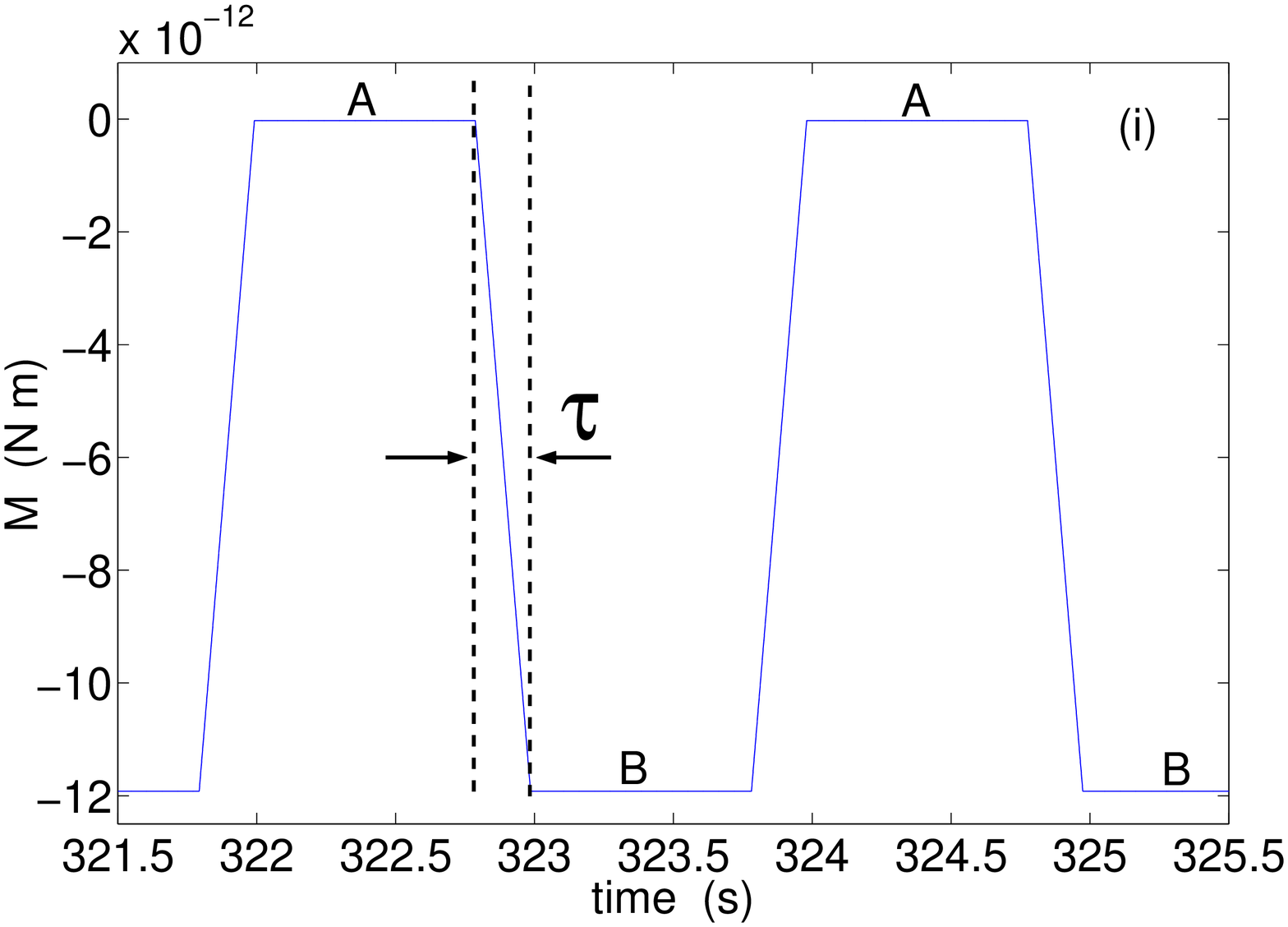} 
    \includegraphics[width=4.6cm, angle=0]{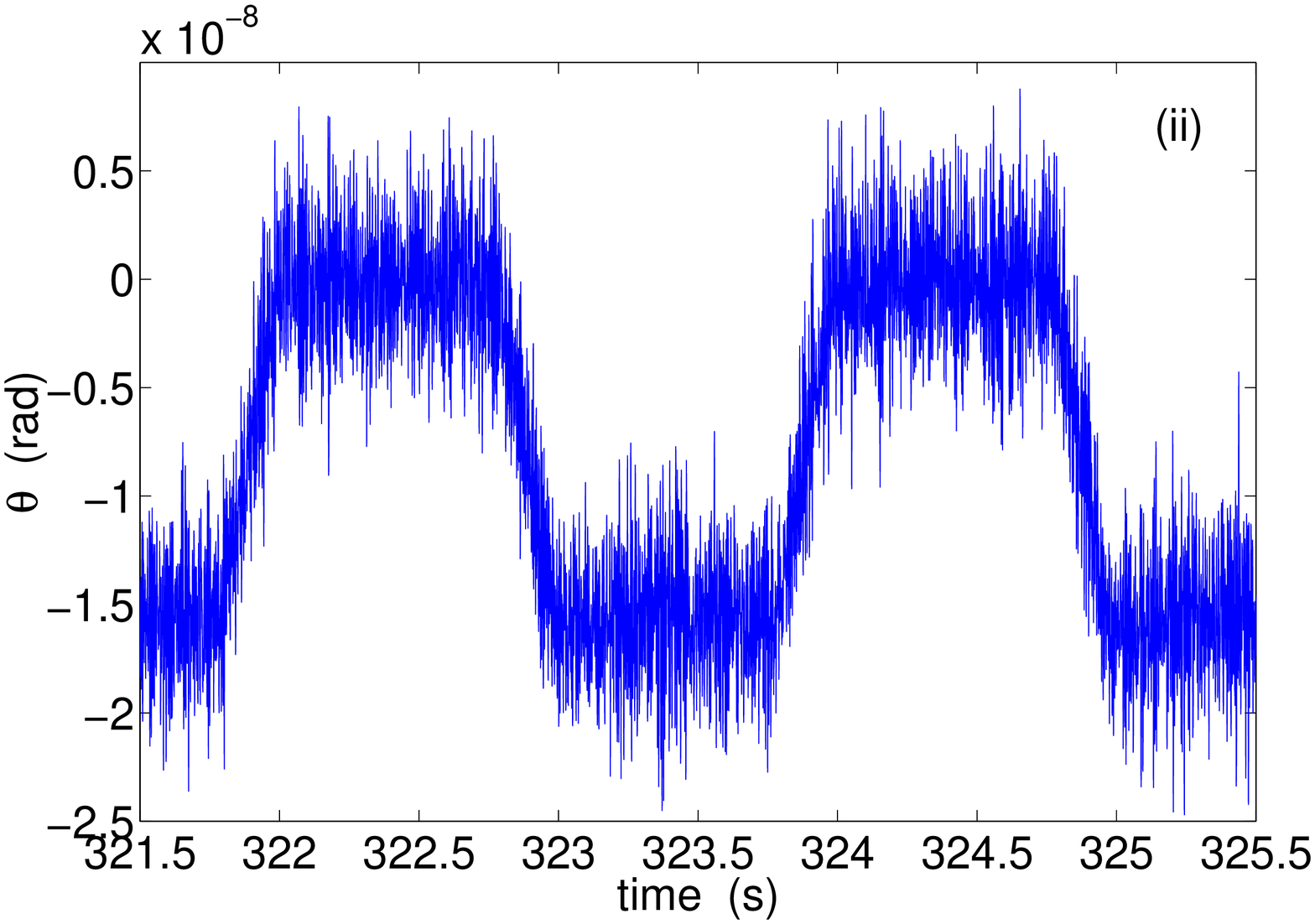}
    \includegraphics[width=4.6cm, angle=0]{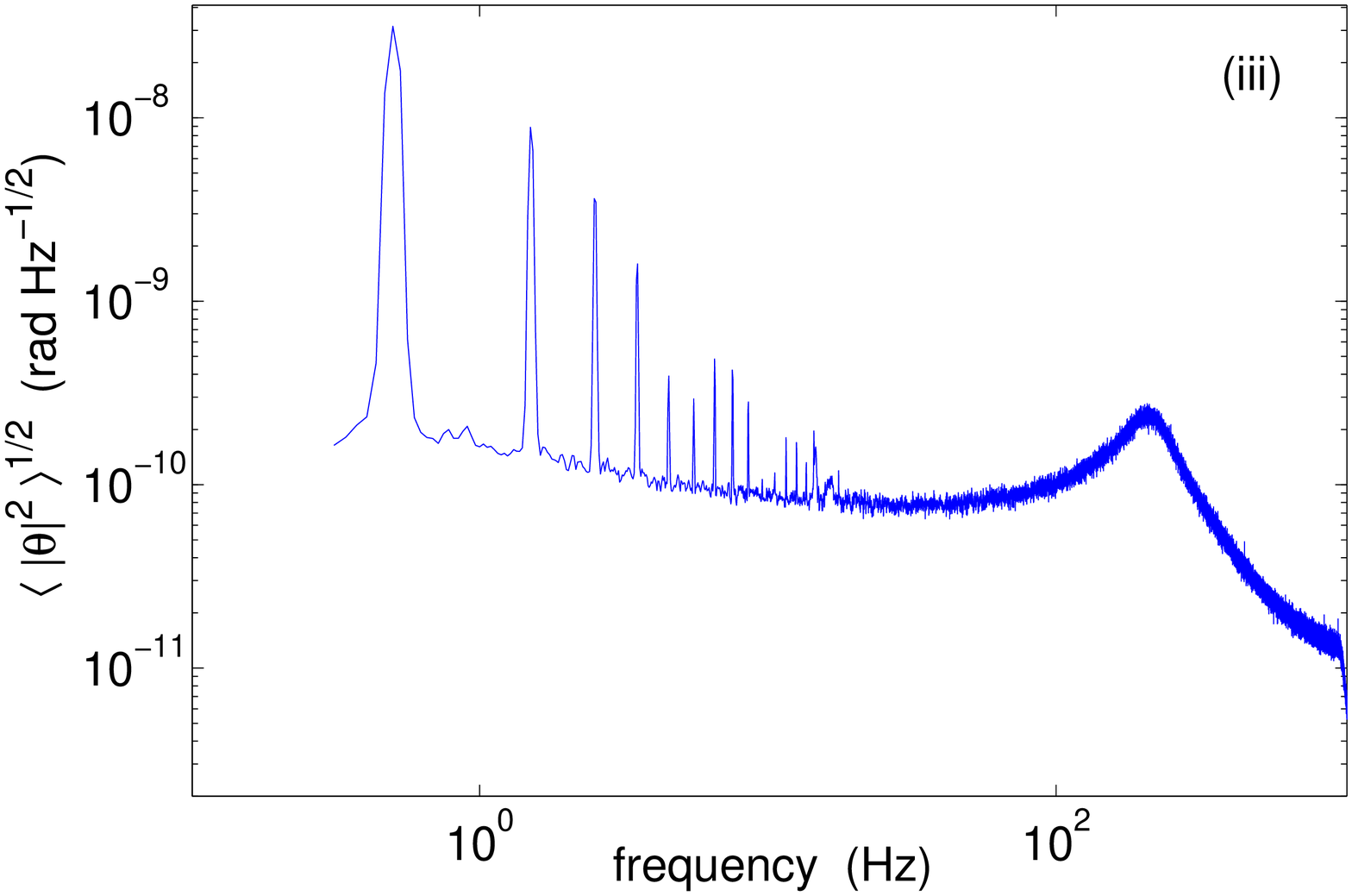}\\
    \includegraphics[width=4.6cm, angle=0]{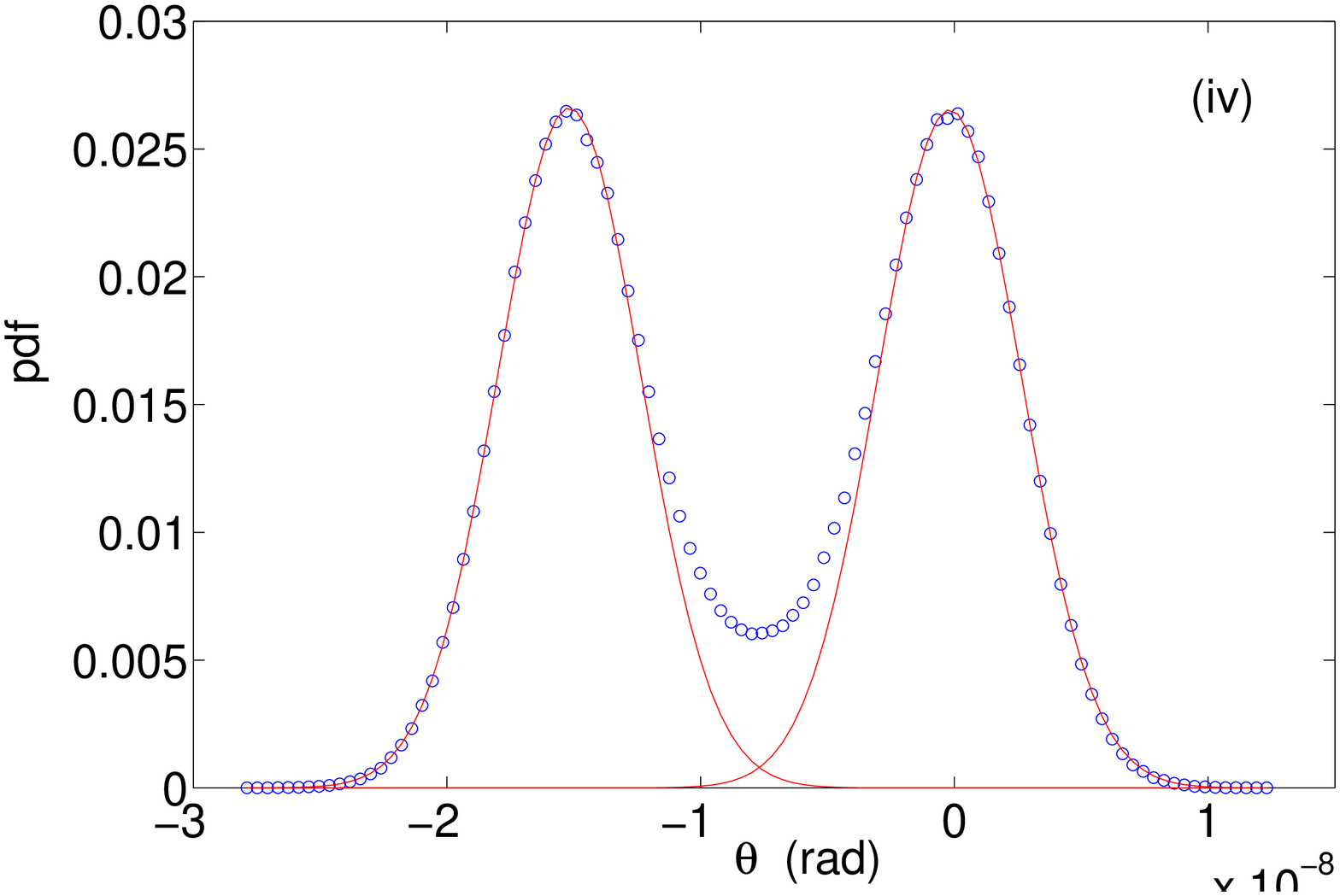}
    \includegraphics[width=4.6cm, angle=0]{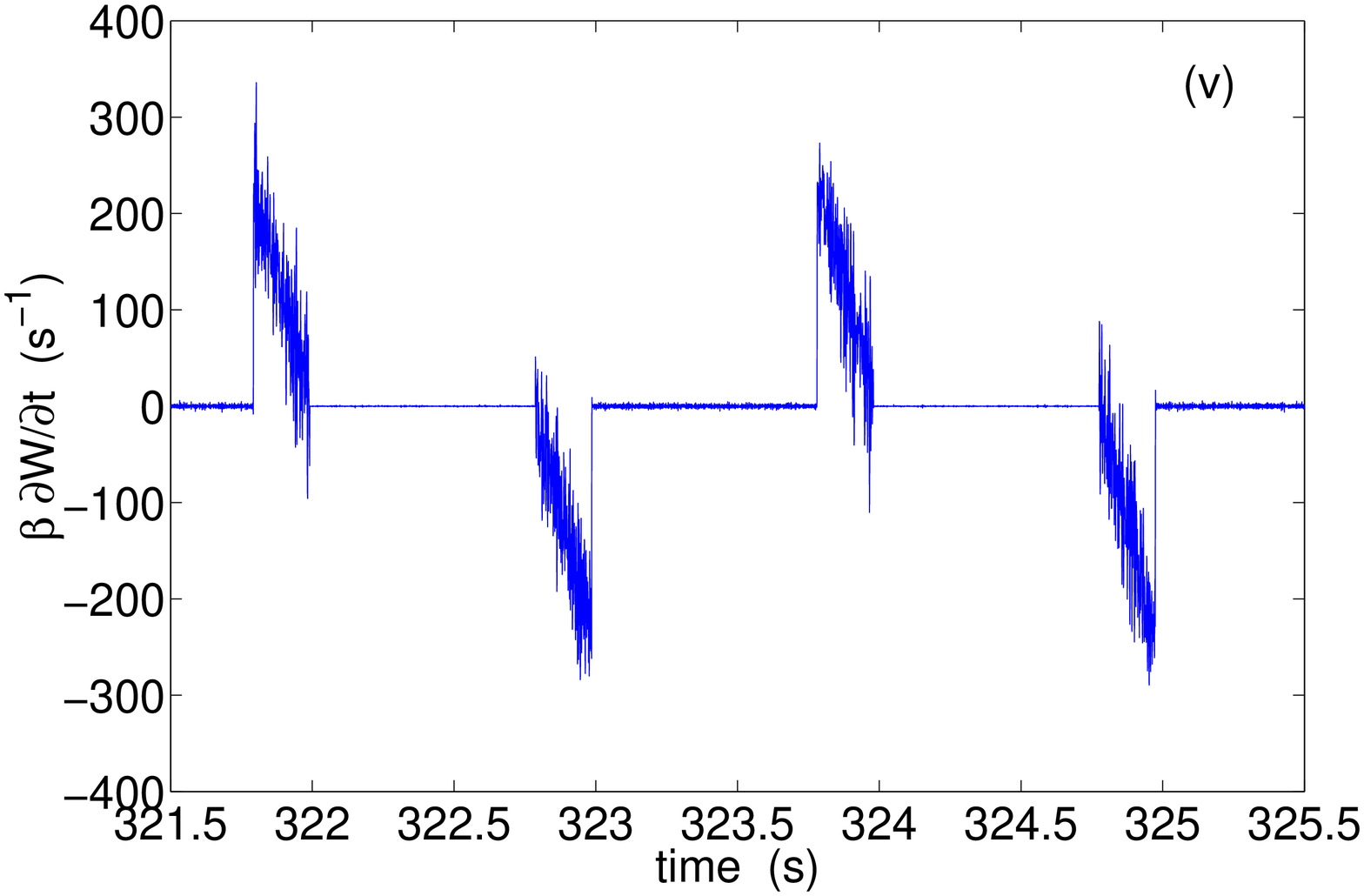}
    \includegraphics[width=4.6cm, angle=0]{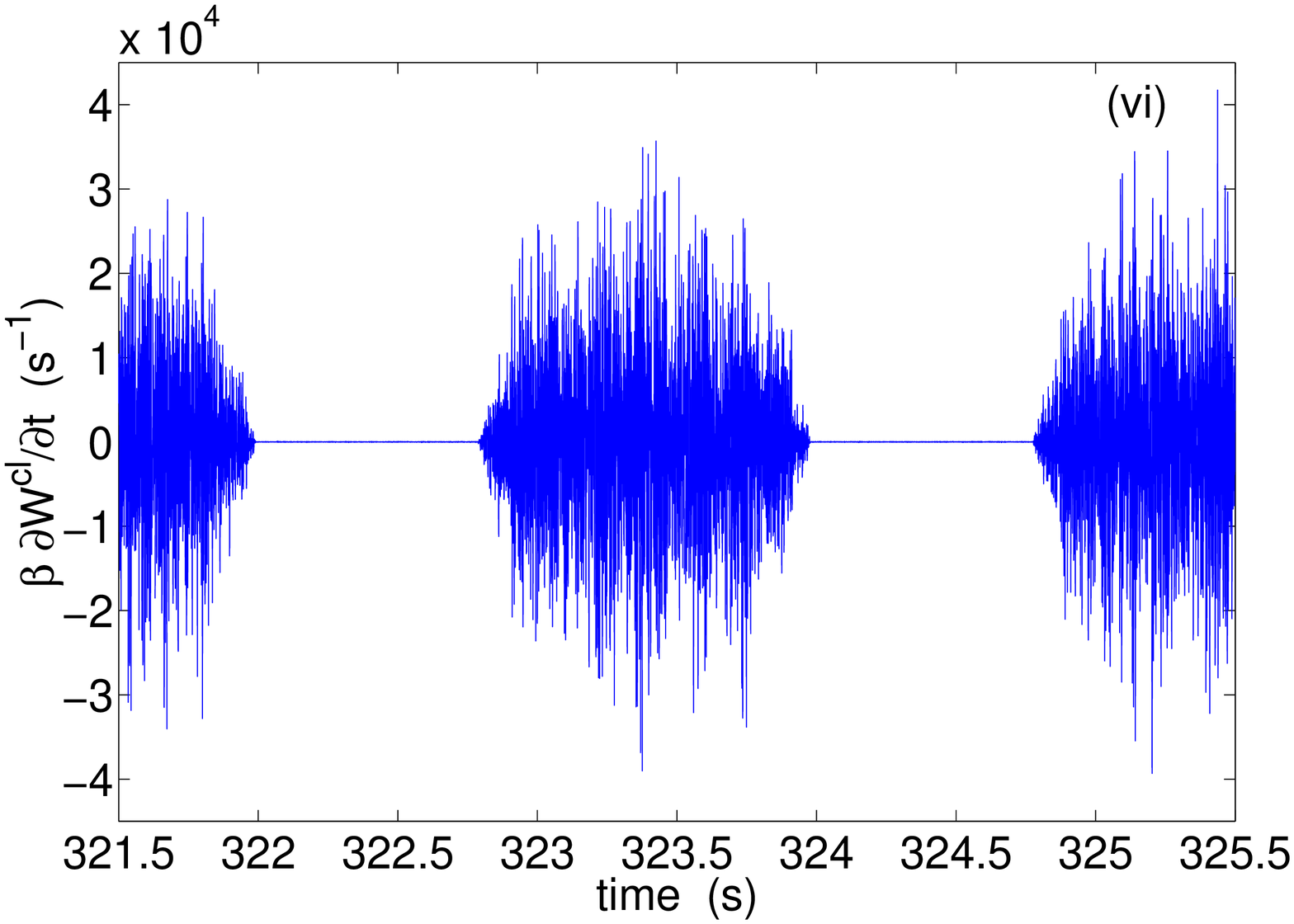}
    \caption{
    Oscillator immersed in oil [case (a)]:
    (i) Applied external torque,
    (ii) Induced angular displacement,
    (iii) its psd, 
    (iv) its pdf,
    (v) Injected power computed from the Jarzynski definition $\dot{W} = -\dot{M} \theta$,
    (vi) Injected power computed from the standard definition $\dot{W}^{\mathrm{cl}} = -M \dot{\theta}$}
    \label{driver_fluct_spec_displacement_displacement_pdf_power_class_power}
    \end{center}
\end{figure}
\section{Experimental setup}
To study the JE and the CR we measure the out-of-equilibrium fluctuations of a
macroscopic mechanical torsion pendulum made of a brass wire, whose damping is
given either by the viscoelasticity of the torsion wire or by the viscosity of a
surrounding fluid. This system is enclosed in a cell which can be filled with a
viscous fluid, which acts as a heat bath. A brass wire of length
$10 \textrm{ mm}$, width $0.75 \textrm{ mm}$, thickness
$50 \textrm{ } \mu \textrm{m}$, mass $5.91 \times 10^{-3} \textrm{ g}$, is
clamped at both ends, hence its elastic torsional stiffness is
$C = 7.50 \times 10^{-4} \textrm{ N\,m}\mathrm{\,rad^{-1}}$. A small mirror of
effective mass $4.02 \times 10^{-2} \textrm{ g}$, length $2.25 \textrm{ mm}$,
width $b_1 = 7 \textrm{ mm}$, thickness $a_1 = 1.04 \textrm{ mm}$, is glued in
the middle of the wire, see Fig.\,\ref{setup_and_fdt_check}(i), so that the
moment of inertia of the wire plus the mirror in vacuum is
$I = 1.79 \times 10^{-10} \textrm{ kg}\mathrm{\,m^2}$ (whose main contribution
comes from the mirror). Thus the resonant frequency of the pendulum in vacuum is
$f_0 = 326.25 \textrm{ Hz}$. When the cell is filled with a viscous fluid, the
total moment of inertia is $I_{\mathrm{eff}} = I + I_{\mathrm{fluid}}$, where
$I_{\mathrm{fluid}}$ is the extra moment of inertia given by the fluid displaced
by the mirror \cite{lamb}. Specifically, for the oil used in the experiment
(which is a mineral oil of optical indice $n = 1.65$, viscosity
$\nu = 121.3 \textrm{ mPa\,s}$ and density
$\rho = 0.9\,\rho_{\mathrm{water}}$ at $T = 21.3 \,^{\circ}\mathrm{C}$)
the resonant frequency becomes $f_0 = 213 \textrm{ Hz}$. To apply an external
torque $M$ to the torsion pendulum, a small electric coil connected to the brass
wire is glued in the back of the mirror. Two fixed magnets on the cell facing
each other with opposite poles generate a static magnetic field. We apply a
torque by varying a very small current $J$ flowing through the electric coil,
hence  $M \propto J$.
The measurement of the angular displacement of the mirror $\theta$ is done using
a Nomarski interferometer \cite{nomarski, optics_com} whose noise is about
$6.25 \times 10^{-12} \textrm{ rad}/\sqrt{\textrm{Hz}}$, which is two orders of
magnitude smaller than the oscillator thermal fluctuations. A window on the cell
allows the laser beams to go inside and outside. Much care has been taken in
order to isolate the apparatus from the external mechanical and acoustic noise,
see \cite{rsi} for details.

The motion of the torsion pendulum can be assimilated to that of a harmonic
oscillator damped by the viscoelasticity of the torsion wire and the viscosity
of the surrounding fluid, whose motion equation reads in the temporal domain
\begin{equation}
    I_{\mathrm{eff}}\,\ddot{\theta} + \int_{-\infty}^{t} G(t-t')\, \dot{\theta}(t') \d t' + C \theta = M,
\end{equation}
where $G$ is the memory kernel. In Fourier space (in the frequency range of our
interest) this equation takes the simple form
$[- I_{\mathrm{eff}}\,{\om}^2 + \hat{C}]\, \hat{\theta} = \hat{M}$,
where $\hat{\cdot}$ denotes the Fourier transform and
$\hat{C} = C + i [C_1'' + \om C_2'']$ is the complex frequency-dependent elastic
stiffness of the system. $C_1''$ and $C_2''$ are  the viscoelastic and viscous
components of the damping term. The response function of the system
$\hat{\chi} = \hat{\theta} / \hat{M}$ can be measured by applying a torque with
a white spectrum. When $M = 0$, the amplitude of the thermal vibrations of the
oscillator is related to its response function via the fluctuation-dissipation
theorem (FDT) \cite{landau_stat}. Therefore, the thermal fluctuation power
spectral density (psd) of the torsion pendulum reads for positive frequencies
\begin{equation}
    \langle { \vert \hat{\theta} \vert }^2 \rangle
    = \frac{4 k_B T}{\om} \, \mathrm{Im} \, \hat{\chi}
    =\frac{4 k_B T}{\om} \frac{C_1'' + \om \, C_2''}
    {{\lbrack -I_{\mathrm{eff}}\,{\om}^2 + C \rbrack}^2 + [C_1'' + \om \, C_2'']^2}.
    \label{fdt}
\end{equation}
We plot in Fig.\,\ref{setup_and_fdt_check}(ii) the measured thermal square root
psd of the oscillator. The measured noise spectrum [circles in
Fig.\,\ref{setup_and_fdt_check}(ii)] is compared with the one estimated [dotted
line in Fig.\,\ref{setup_and_fdt_check}(ii)] by inserting the measured
$\hat{\chi}$ in the FDT, Eq.\,(\ref{fdt}). The two measurements are in perfect
agreement and obviously the FDT is fully satisfied because the system is at
equilibrium in the state $A$ where $M = 0$ (see below). Although this result is
expected, this test is very useful to show that the experimental apparatus can
measure with a good accuracy and resolution the thermal noise of the macroscopic
pendulum.
\begin{figure}
    \begin{center}
    \includegraphics[width=6cm, angle=0]{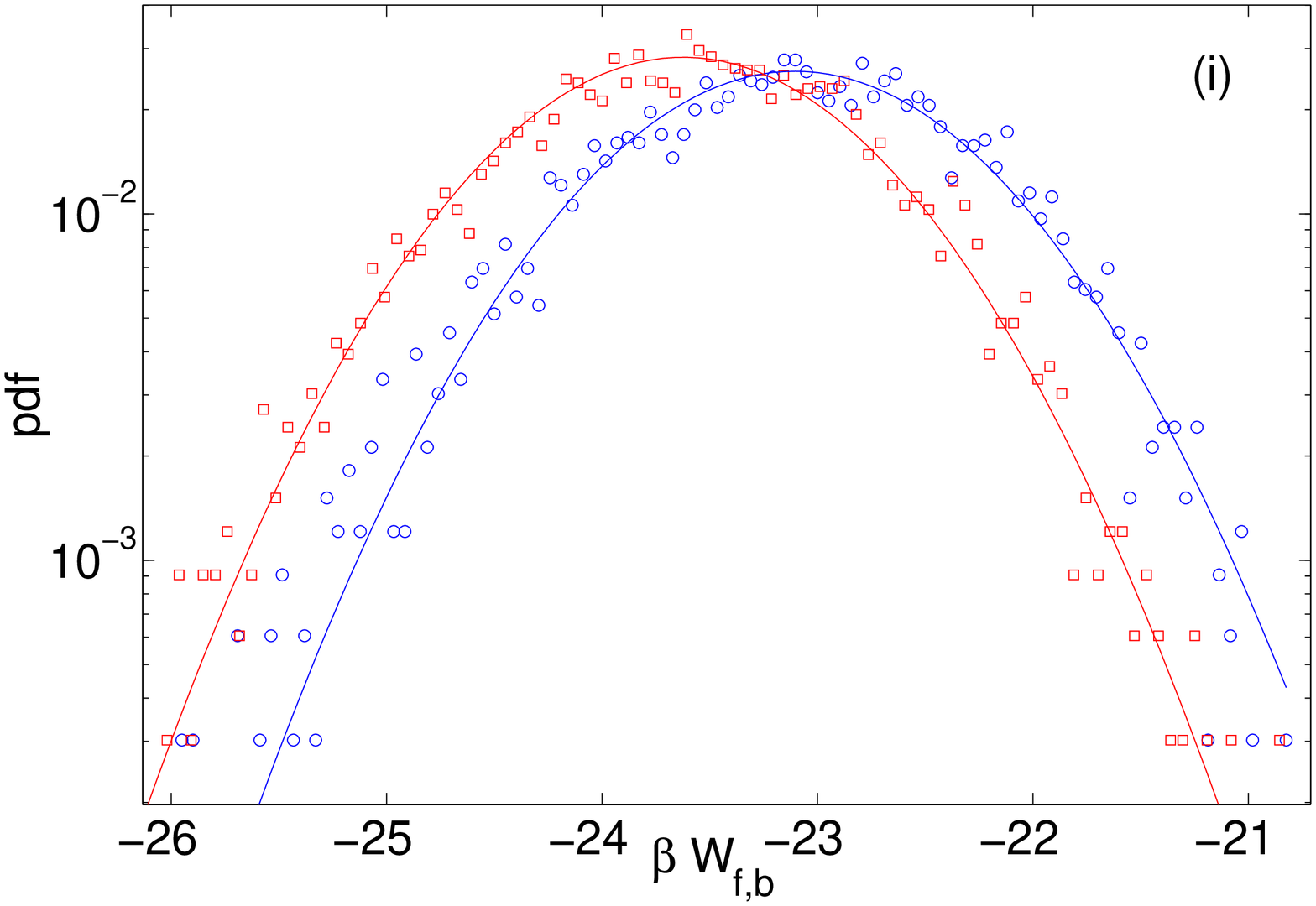}
    \includegraphics[width=6cm, angle=0]{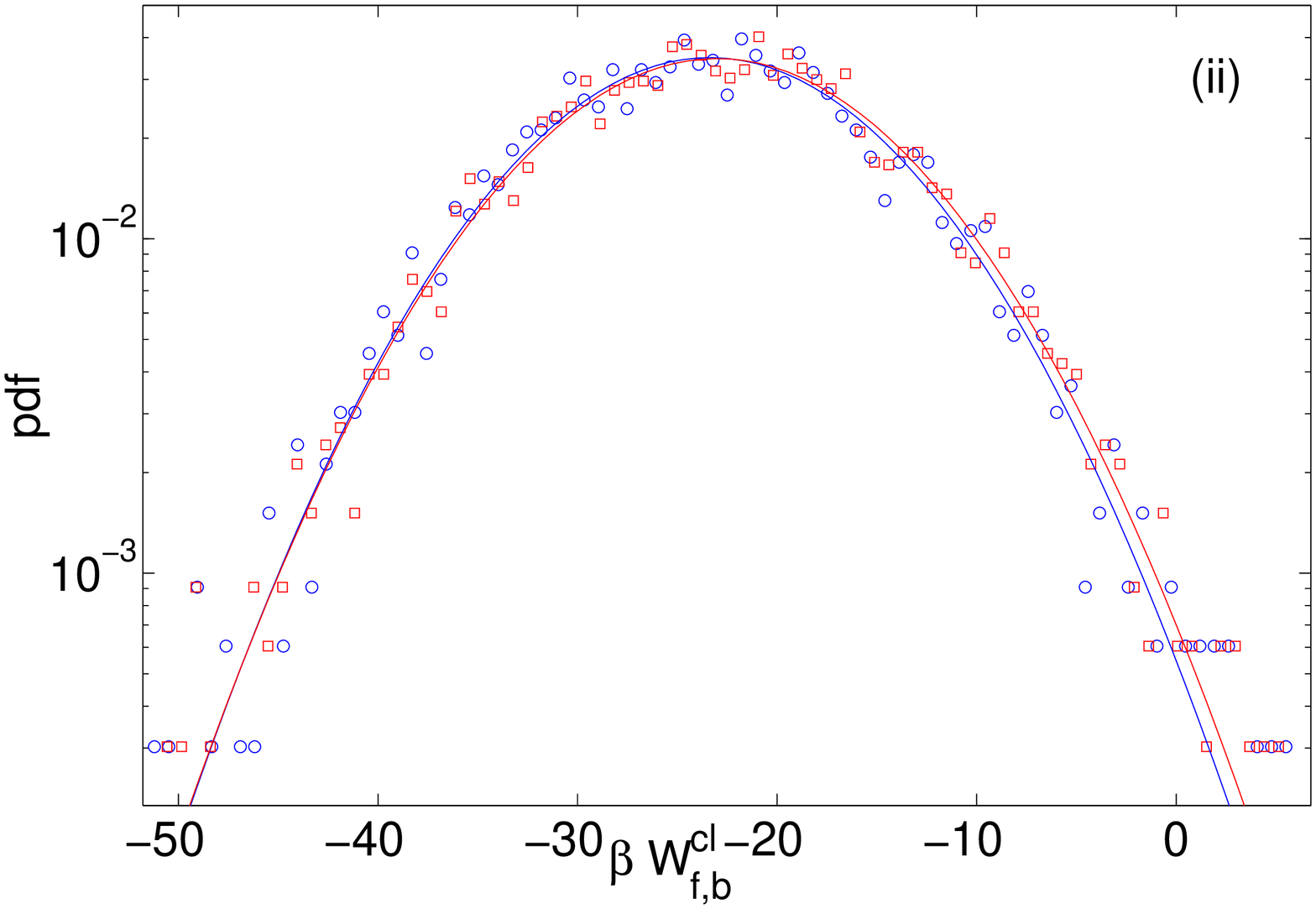}\\
    \includegraphics[width=6cm, angle=0]{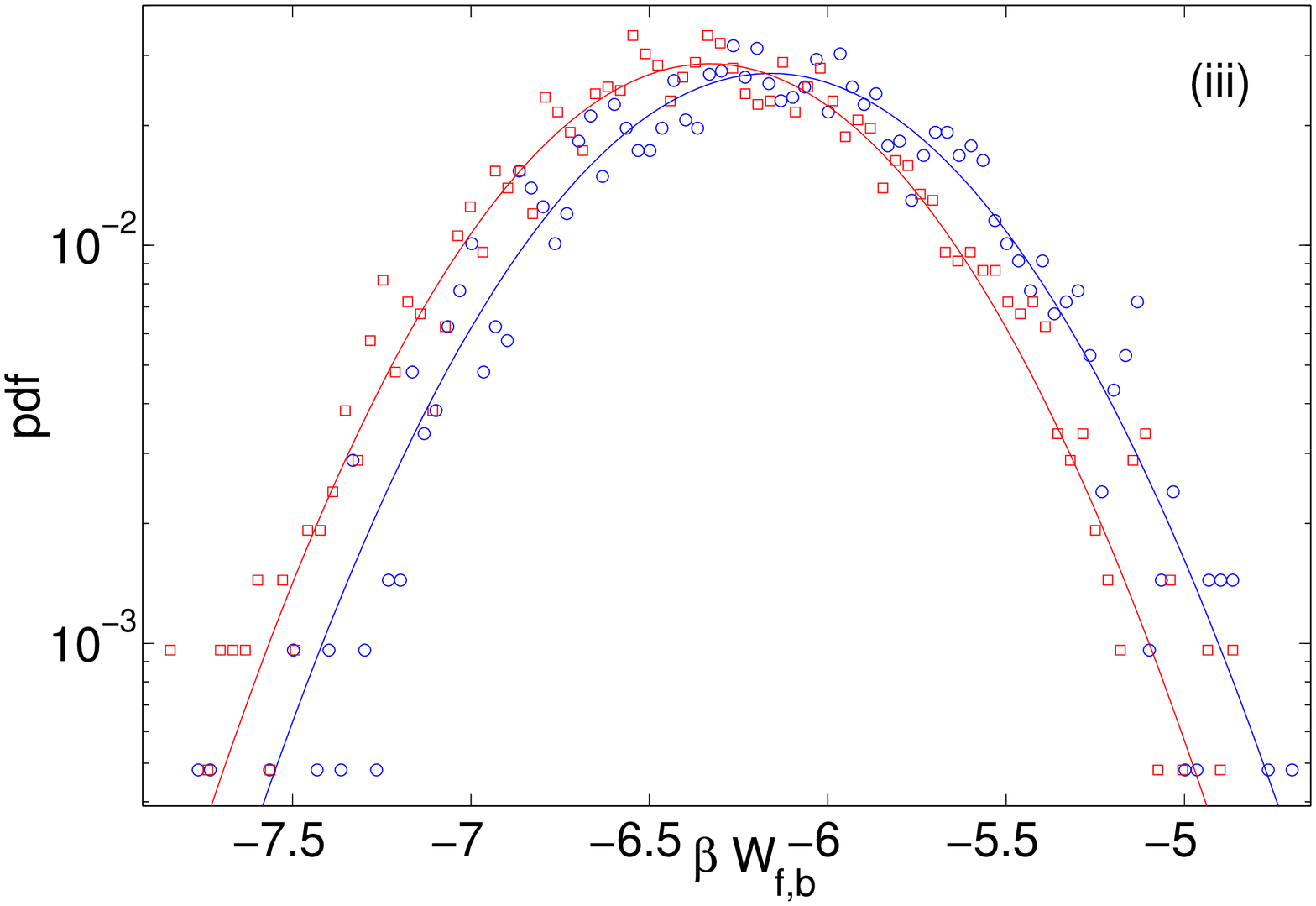}
    \includegraphics[width=6cm, angle=0]{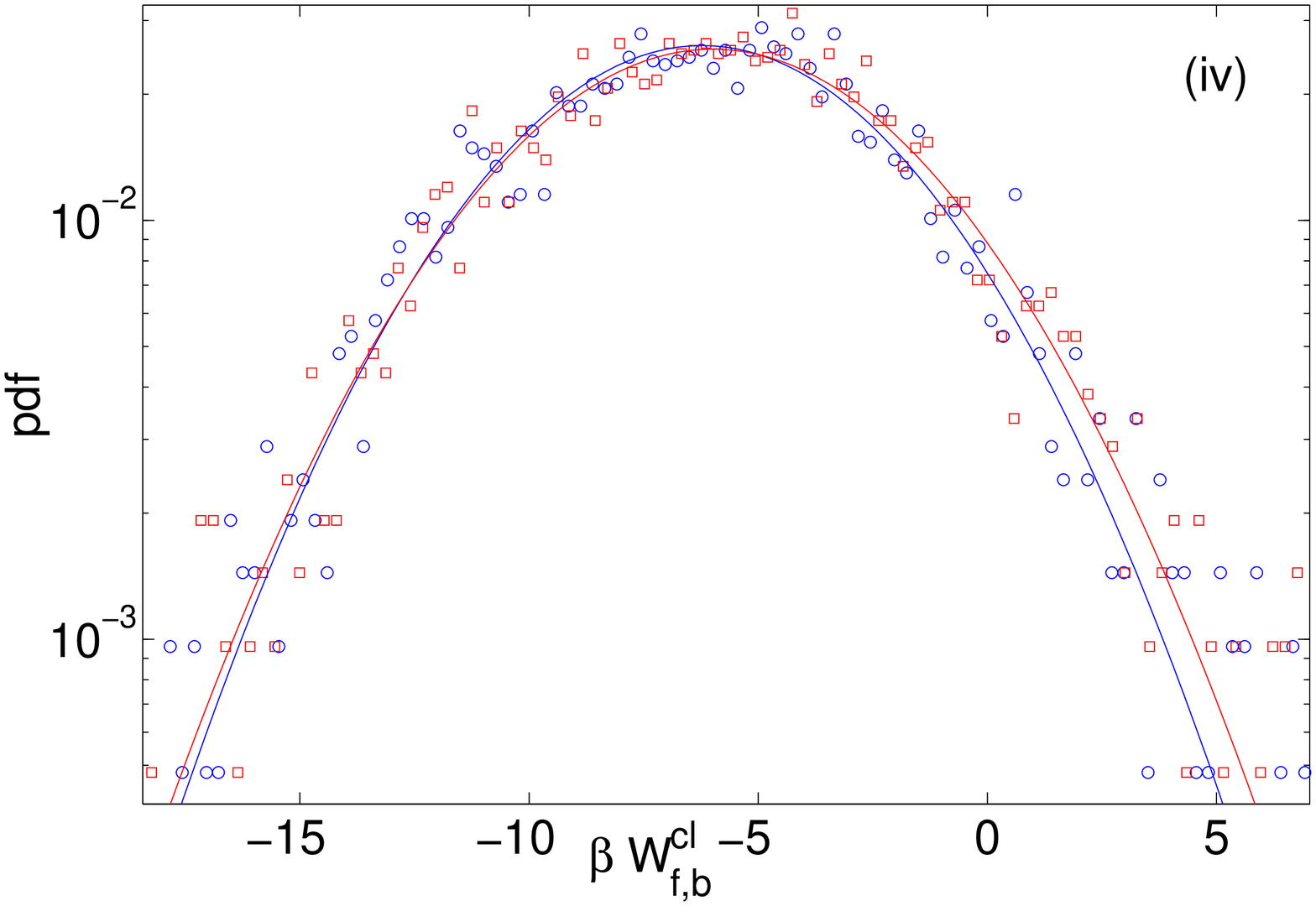}
    \caption{Case a): (i) $\mathrm{P}_{\mathrm{f}}(W)$ and $\mathrm{P}_{\mathrm{b}}(-W)$,
    (ii) $\mathrm{P}_{\mathrm{f}}(W^{\mathrm{cl}})$ and $\mathrm{P}_{\mathrm{f}}(-W^{\mathrm{cl}})$;
    Case c): (iii) $\mathrm{P}_{\mathrm{f}}(W)$ and $\mathrm{P}_{\mathrm{b}}(-W)$,
    (iv) $\mathrm{P}_{\mathrm{f}}(W^{\mathrm{cl}})$ and $\mathrm{P}_{\mathrm{f}}(-W^{\mathrm{cl}})$
    (experimental forward and backward pdfs are represented by $\circ$ and $\Box$ respectively,
    whereas the continuous lines are Gaussian fits)}
    \label{pdfs}
    \end{center}
\end{figure}
\section{Experimental results} Now we drive the oscillator out of equilibrium
between two states $A$ (where $M = 0$) and $B$ (where
$M = M_{\mathrm{max}} = \mathrm{const} \neq 0$). The path $\ga$ may be changed
by modifying the time evolution of $M$ between $A$ and $B$. We have chosen
either linear ramps with different rising times $\tau$, see
Fig.\,\ref{driver_fluct_spec_displacement_displacement_pdf_power_class_power}(i),
or half-sinusoids with half-period $\tau$. In the specific case of our harmonic
oscillator, as the temperature is the same in states $A$ and $B$, the free
energy difference of the oscillator alone is
$\Delta F_0 = \Delta U = \Big[ \frac{1}{2} C \theta^2 \Big]_A^B = \Big[ \frac{M^2}{2C} \Big]_A^B$,
whereas $\Delta F = \Delta F_0 - \Big[ \frac{M^2}{C} \Big]_A^B$, i.e. for an
harmonic potential $\Delta F = -\Delta F_0$. Let us first consider the situation
where the cell is filled with oil. The oscillator's relaxation time is
$\tau_{\mathrm{relax}} = 23.5 \textrm{ ms}$. We apply a torque which is a
sequence of linear increasing\,/decreasing ramps and plateaux, as represented in
Fig.\,\ref{driver_fluct_spec_displacement_displacement_pdf_power_class_power}(i).
We chose different values of the amplitude of the torque $M$
[$11.9$, $6.1$, $4.2$ and $1.2$ pN\,m] and of the rising time $\tau$
[$199.5$, $20.2$, $65.6$, $99.6$ ms, respectively], as indicated in
Table \ref{results} [cases a)\dots e)]. Thus we can probe either the reversible
(or quasi-static) paths ($\tau \gg \tau_{\mathrm{relax}}$) or the irreversible
ones ($\tau \lesssim \tau_{\mathrm{relax}}$). We tune the duration of the
plateaux (which is at least $4\, \tau_{\mathrm{relax}}$) so that the system
always reaches equilibrium in the middle of each of them, which defines the
equilibrium states $A$ and $B$. We see in
Fig.\,\ref{driver_fluct_spec_displacement_displacement_pdf_power_class_power}(ii),
where the angular displacement $\theta$ is plotted as a function of time
[case a)], that the response of the oscillator to the applied torque is
comparable to the thermal noise spectrum. The psd of $\theta$ is shown in
Fig.\,\ref{driver_fluct_spec_displacement_displacement_pdf_power_class_power}(iii).
Comparing this measure with the FDT prediction obtained in
Fig.\,\ref{setup_and_fdt_check}(ii), one observes that the driver does not
affect the thermal noise spectrum which remains equal to the equilibrium one.
Moreover we plot in
Fig.\,\ref{driver_fluct_spec_displacement_displacement_pdf_power_class_power}(iv)
the pdf of the driven displacement $\theta$ shown on
Fig.\,\ref{driver_fluct_spec_displacement_displacement_pdf_power_class_power}(ii),
which is, roughly speaking, the superposition of two Gaussian pdfs. From the
measure of $M$ and $\theta$, the power injected into the system $\dot{W}$ can be
computed from the definition given in Eq.\,(\ref{work}), that in this case is
$\dot{W} = -\dot{M} \theta$. Its time evolution, shown in
Fig.\,\ref{driver_fluct_spec_displacement_displacement_pdf_power_class_power}(v),
is quite different from that of the classical power
$\dot{W}^{\mathrm{cl}} = -M \dot{\theta}$, whose time evolution is
plotted on
Fig.\,\ref{driver_fluct_spec_displacement_displacement_pdf_power_class_power}(vi):
$\dot{W}$ is non-zero only for $\dot M \neq 0$ and vice-versa
$\dot{W}^{\mathrm{cl}} \neq 0$ only for $M \neq 0$. From one time series of
$\dot{W}$ we can compute from Eq.\,(\ref{work}) the forward and the backward
works, $W_\mathrm{f}$ and $W_\mathrm{b}$, corresponding to the paths $A \to B$
and $B \to A$, respectively. We also do the same for the classical work. We then
compute their respective pdfs $\mathrm{P}_{\mathrm{f}}(W)$ and
$\mathrm{P}_{\mathrm{b}}(-W)$. These are plotted on Figs.\,\ref{pdfs}(i,iv)
where the bullets are the experimental data and the continuous lines their
fitted Gaussian pdfs. In Figs.\,\ref{pdfs}, the pdfs of $W$ and
$W^{\mathrm{cl}}$ cross in the case a) at $\beta W \simeq -23.5$, and in the
case c) $\beta W \simeq -6.1$. These values correspond to the
$\Delta F = -\frac{M_{\mathrm{max}}^2}{2C} = -\Delta F_0$. We find that this
result is true independently of the ratio $\tau / \tau_{\mathrm{relax}}$ and of
the maximum amplitude of $|M|$, $M_{\mathrm{max}}$. This can be seen on Table
\ref{results}, where the computed $\Delta U = {M_{\mathrm{max}}^2 \over 2C}$
is in good agreement with the values obtained by the crossing points of the
forward and backward pdfs, that is $\Delta F_{\times} + \Phi$ for
$\mathrm{P}(W)$ and  $-\Delta W_{\times}^{\mathrm{cl}}$ for
$\mathrm{P}(W^{\mathrm{cl}})$. Finally inserting the values of $W_{\mathrm{f}}$
and $W_{\mathrm{b}}$ in Eq.\,(\ref{JE}) we directly compute
$\Delta F_{\mathrm{f}}$ and $\Delta F_{\mathrm{b}}$ from the JE. As can be seen
in Table \ref{results}, the values of $\Delta F_0$ obtained from the JE, that is
either $-(\Delta F_{\mathrm{f}} + \Phi)$ or $-(\Delta F_{\mathrm{b}} + \Phi)$,
agree within experimental errors with the computed $\Delta U$. Indeed JE works
well either in the foreseeable case a) where $\tau \gg \tau_{\mathrm{relax}}$
or in the critical case b) where $\tau \lesssim \tau_{\mathrm{relax}}$.
The other case we have studied is a very pathological one. Specifically, the
oscillator is in vacuum and has a resonant frequency $f_0 = 353 \textrm{ Hz}$
and a relaxation time $\tau_{\mathrm{relax}} = 666.7 \textrm{ ms}$. We applied a
sinusoidal torque whose amplitude is either $5.9 \times 10^{-12}$ or
$9.4 \times10^{-12} \textrm{ N\,m [cases f) and g) in Table \ref{results},
respectively]}$. Half a period of the sinusoid is $\tau = 49.5 \textrm{ ms}$,
much smaller than the relaxation time, so that we never let the system
equilibrate. However, we define the states $A$ and $B$ as the maxima and minima
of the driver. Surprisingly, despite of the pathological definition of the
equilibrium states $A$ and $B$, the pdfs are Gaussian and the JE is satisfied as
indicated in Table \ref{results}. Moreover, this happens independently of
$M_{\mathrm{max}}$ and of the critical value of the ratio
$\tau / \tau_{\mathrm{relax}} \ll 1$. Finally, we indicated in Table
\ref{results} the value $\Delta F_{\circlearrowleft}$ which is the free energy
computed from the JE if one considers the ``loop process'' from $A$ to $A$ (the
same can be done from $B$ to $B$ and the results are quantitatively the same).
In principle this value should be zero, but in fact it is not since we have
about 5\% error in the calibration of the torque $M$.
\begin{table}[!t]
    \begin{center}
    \begin{tabular}{|c|c|c|c|c|c|c|c|}
    \hline
    $\tau / \tau_{\mathrm{relax}}$ & $M_{\mathrm{max}}$ & $-\beta[\Delta F_{\mathrm{f}} + \Phi]$ & $\beta[\Delta F_{\mathrm{b}} + \Phi]$ & $-\beta[\Delta F_{\times} + \Phi]$ & $-\beta W_{\times}^{\mathrm{cl}}$ & $\beta \Delta U$ & $ |\beta \Delta F_{\circlearrowleft}|$ \\
    \hline \hline
    $8.5^{\textrm{ a)}}$ & $11.9$ & $23.5$ & $23.1$ & $23.5$ & $23.4$ & $23.8$ & $1.0$\\
    $0.85^{\textrm{ b)}}$ & $6.1$ & $5.9$ & $5.4$ & $6.0$ & $7.0$ & $6.0$ & $1.0$\\
    $3.5^{\textrm{ c)}}$ & $6.1$ & $6.1$ & $5.9$ & $6.5$ & $6.1$ & $6.1$ & $0.4$\\
    $2.8^{\textrm{ d)}}$ & $4.2$ & $2.8$ & $2.6$ & $3.2$ & $2.9$ & $2.7$ & $0.3$\\
    $4.2^{\textrm{ e)}}$ & $1.2$ & $0.21$ & $0.20$ & $ 0.22 $ & $ 0.21 $ & $0.22$ & $0.04$\\
    \hline \hline
    $0.07^{\textrm{ f)}}$ & $5.9$ & $10.3$ & $10.0$ & $10.1$ & $10.1$ & $10.3$ & $0.4$\\
    $0.07^{\textrm{ g)}}$ & $9.4$ & $67.4$ & $65.5$ & $66.8$ & $66.4$ & $67.5$ & $2.4$\\
    \hline
    \end{tabular}
    \vspace{0.5cm}
    \caption{Free energies of cases a)\dots g) defined in the text (the values of $M_{\mathrm{max}}$ are in pN\ m).
    $\Delta U={M_{\mathrm{max}}^2 \over 2C}$ is the computed expected value}
    \label{results}
    \end{center}
\end{table}
\section{Conclusion}
Our results clearly demonstrate the validity and the robustness of the JE in an
isothermal process, at least when the work fluctuations are Gaussian and when
the harmonic approximation is relevant for the system. The more accurate and
reliable $\Delta F$ estimator is given by the crossing points
$\Delta F_{\times}$ and $W_{\times}^{\mathrm{cl}}$, because they are less
sensitive to extreme fluctuations which may perturb the convergence of the JE.
We have also shown that, in the case of Gaussian fluctuations,
$W_{\times}^{\mathrm{cl}}$ remains an excellent estimator even in cases where
the JE and the CR could not hold. Unfortunately our results do not fully throw
light on the debate between Cohen, Mauzerall and Jarzynski
\cite{cohen, jarzynski4} since the work pdfs are Gaussian. Recently, Ritort and
coworkers have used the JE to estimate $\Delta F$ in an experiment of RNA
stretching where the oscillator's coupling is non-linear and the work
fluctuations are non-Gaussian \cite{ritort_told_us}. It would be interesting to
check these results on a more simple and controlled system. We are currently
working on the experimental realization of such a non-linear coupling, for which
$\Delta F \neq -\Delta F_0$. Finally we want to stress that our results,
although limited to the Gaussian case, show that it is possible to measure tiny
fluctuations of work in a macroscopic systems. As consequence they open a lot of
perspective to use the JE, the CR and the recent theorems on dissipated work
(see for example \cite{cvz}) to characterize the slow relaxation towards
equilibrium in more complex systems, for example aging materials such as glasses
or gels \cite{crisanti}.

The authors thank L. Bellon, E.G.D. Cohen, N. Garnier, C.
Jarzynski, F. Ritort and L. Rondoni for useful discussions, and
acknowledge P. Metz, M. Moulin, P.-E. Roche and F. Vittoz for
technical support. This work has been partially supported by the
{\sc{Dyglagemem}} contract of EEC.
\end{document}